# A Safer, Smaller, Cleaner Subcritical Thorium Fission - Deuteron Fusion Hybrid Reactor: DD Collider Instead of Muonic Fusion

D. Akturk,[a*] A. C. Canbay,[b] B. Dagli,[a] U. Kaya,[b] and S. Sultansoy[a,c]

[a]TOBB University of Economics and Technology, Ankara, Türkiye
[b]Ankara University, Ankara, Türkiye
[c]ANAS Institute of Physics, Baku, Azerbaijan

[*]Corresponding Author: dilaraakturk@etu.edu.tr

**Abstract**

Fossil fuels, which meet most of humanity's energy needs, cause climate change due to their high carbon emissions. There are two types of energy sources that can replace fossil fuels: renewable and nuclear. Nuclear energy sources are more advantageous in terms of efficiency and sustainability. The use of Thorium as nuclear fuel in fusion reactors will contribute to the reduction of radioactive waste, due to the much lower production of transuranics. Fusion reactors, which are considered promising, are still in the R&D phase. In this respect, hybrid fusion-fission reactors seem more promising and the recently proposed combination of muon-catalyzed *DD* fusion with a cascade thorium reactor is worthy of appreciation. In this study, we show that using the *DD* collider instead of muonic fusion has significant advantages.

**Keywords:** *DD* Collider, Thorium, Hybrid Reactors, Fusion, Fission, Nuclear Energy

## 1. Introduction

Given global warming, it is imperative that humanity should turn to different energy sources as an alternative to fossil fuels. Today there are two options for replacing fossil energy sources: renewable energy (solar, wind, geothermal, hydro, etc.) and nuclear energy. Nuclear reactors can make an exceptional contribution in this field, because of their efficiency and sustainability. At COP29 UN Climate Conference, it was proposed to triple the generation of electricity from nuclear power over the next 20 years [1].

There are three possible options for nuclear power generation: fission, fusion and hybrid reactors. Concerning fission reactors, the most important problem is radioactive waste. This problem can be much more tolerable when Thorium fuel is used [2].

As for fusion, TOKAMAK related studies have been conducted in this area for over 70 years and the ITER project was launched 20 years ago in this context. It started with 5 billion euros and was planned to be operational in 2015, but now the budget has exceeded 25 billion euros, and the operation has been postponed to 2035. Regarding laser-driven fusion systems, the 1.5 energy gain reported is based on comparison to the energy of the laser beam. Keeping in mind the efficiency of the laser, the actual gain is significantly lower than 5%.

Regarding accelerator-based Thorium reactors, the most advanced system is Rubbia's Energy Amplifier (EA) project [3] and, in this context, the MYRRHA project [4] was initiated 10 years ago. This project aims to build a demonstration of EA using a 10 mA beam current proton accelerator with 600 MeV energy. The total cost of the project is around 1 billion euros and Belgium has allocated 600 million, the remaining 400 million has not yet been provided by the EU.

We believe that hybrid reactors with Thorium fuel may become more perspective in coming decades. In a recent paper [5], Subcritical Thorium Fission - Muonic Fusion Hybrid Reactor has been proposed. In this study, we suggest the use of a Deuteron collider as a neutron source for Thorium fission, instead of muon-catalyzed fusion.

Muon catalyzed fusion proposal [5] is briefly discussed in Section 2. In Section 3, deuterium collider option has been presented, where calculation of luminosity and deuteron accelerator parameters for deuteron collider are given together with energy balance. Finally, our conclusions and suggestions are presented in Section 4.

## 2. Muon Catalyzed Fusion Version

The schematic view of the hybrid reactor, proposed in [5], is shown in Figure 1. The fusion neutron core (red) is in the center of the cylindrical reactor, while the cascade is made up of thin coaxial cylinders containing highly enriched $^{233}$U. The space between the inner and outer cylinders is filled with light water, inside which highly enriched $^{233}$U rods are arranged in a picket fence–like pattern, illustrated as black dots in the cascade, which is colored green in the figure (for details see [5]).

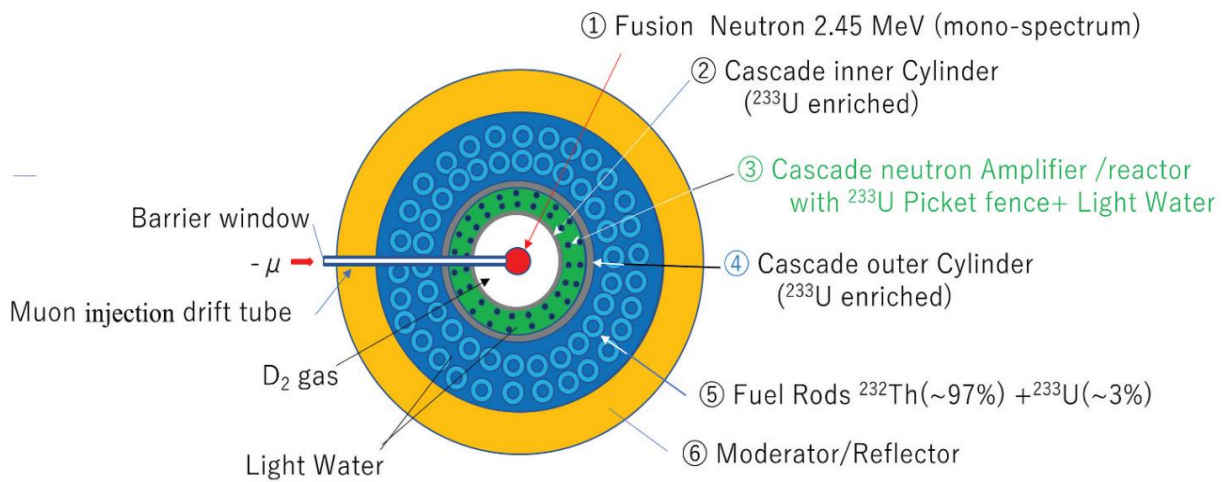

Figure 1. Schematic diagram of the hybrid reactor with cascade booster section (Fig. 2 in [5])

The table of energy balance Q values for $DT\mu$ and $DD\mu$ fusions (Table 1 in [5]) is presented below. Unfortunately, there are some typographical errors in the original version of the table. The corrected values are presented in Table 1 within curly brackets. It should be also mentioned that the electrical power is approximately half of the thermal power. Therefore, 5 MW of electrical power will be generated, instead of 10 MW.

Table 1. Energy Balance Q Values for *DTμ* and *DDμ* Fusions

|  | ① Muon catalyst cycle or cascaded neutron amplification | ② Fission Multiplication $k_{eff}$, M = $1/(1-k_{eff})$ | ③ Muon Numbers (Bq) | ④ Output Electric (MW) | ⑤ Input Electric | ⑥ Qeconomy ④ ÷ ⑤ |
|---|---|---|---|---|---|---|
| *DTμ* | 150 |  | $6.6×10^{16}$ | 10 {5} | 16 MW | 0.625 < 1 |
| *DDμ* | 1 |  | $5.1×10^{19}$ | 10 {5} | 12400 GW {12400 MW} | $8.2×10^{-5}$ {$8.2×10^{-4}$} |
| Cascade1 | M1 = 20 | 0.95 | $2.55×10^{18}$ | 10 {5} | 612 MW | $1.64×10^{-3}$ {$1.64×10^{-2}$} |
| Cascade1× Cascade2 | M1×M2 = 1000 | 0.98, M2 = 50 | $2.55×10^{15}$ {$5.10×10^{16}$} | 10 {5} | 122 MW {12.2 MW} | 0.082 {0.82} |
| Cascade1× Cascade2× main reactor | M1×M2×MR = 100 000 | 0.99, MR = 100 | $2.55×10^{13}$ {$5.10×10^{14}$} | 10 {5} | 1.22 MW {0.12 MW} | 8.2 {82} |

## 3. The Deuterium Collider Fusion Option

In muon-catalyzed version, a proton accelerator with an energy of 600 MeV is required for muon production. In this case, issues related to the accelerator's reliability and sustainability may arise. In contrast, the DD collider version is much simpler in terms of accelerator design: deuteron beams with energy on the order of 100 keV are sufficient.

### 3.1. Parameters of the Deuteron Collider

A schematic representation of the hybrid system based on the DD collider is shown in Figure 2. In this configuration, the inner region contains a vacuum environment, instead of D gas.

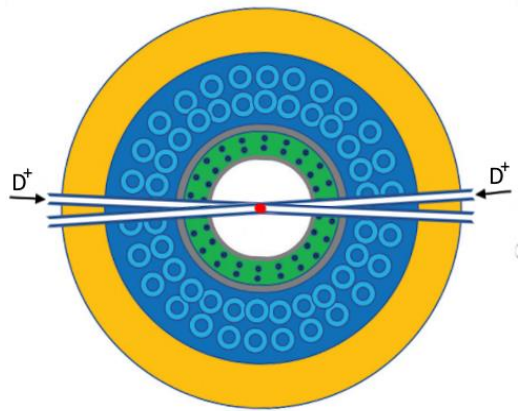

Figure 2. The Schematic Illustration of the DD Collider-Based Hybrid Reactor

Table 2 presents the final states produced in DD and DT collisions. It is seen that 14 MeV neutrons are generated in DT collisions and at these neutron energies, interactions with Thorium result in the production of $^{231}$Pa and $^{231}$Th, leading to the release of strong alpha and gamma radiation (see Fig. 8 in [5]). Therefore, despite the higher fusion cross-section of DT compared to DD fusion (see Figure 3), we will solely examine the DD collider option, for the similar reason as muon-catalyzed fusion version: with neutron energy at 2.4 MeV, the alpha and gamma radiation issue is resolved.

Table 2. Final States Produced in DD and DT Collisions

| | | | | | | | | |
|---|---|---|---|---|---|---|---|---|
| $^{2}_{1}D$ | $+\,^{3}_{1}T$ | $\rightarrow$ | $^{4}_{2}He$ | $(3.52\ MeV)$ | $+\,n^{0}$ | $(14.06\ MeV)$ | | (1) |
| $^{2}_{1}D$ | $+\,^{2}_{1}D$ | $\rightarrow$ | $^{3}_{1}T$ | $(1.01\ MeV)$ | $+\,p^{+}$ | $(3.02\ MeV)$ | %50 | (2i) |
| | | $\rightarrow$ | $^{3}_{2}He$ | $(0.82\ MeV)$ | $+\,n^{0}$ | $(2.45\ MeV)$ | %50 | (2ii) |

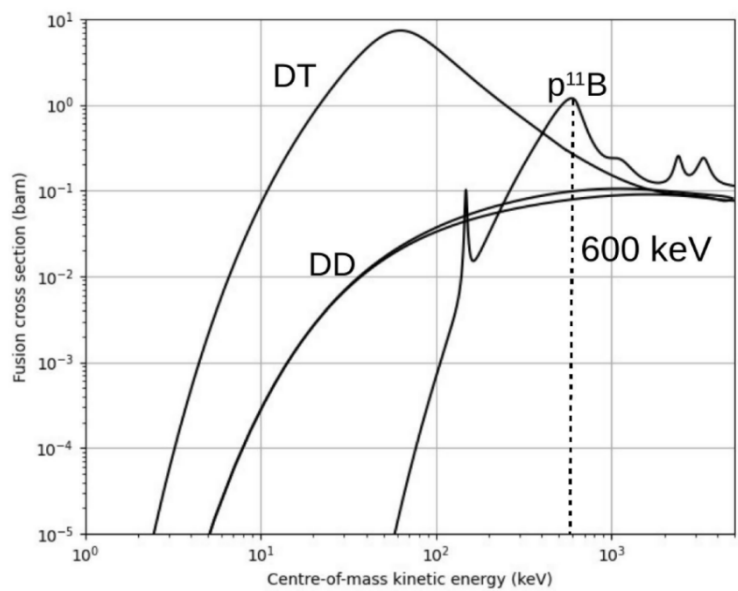

Figure 3. Fusion Cross-Sections [6]

### 3.1.1. Luminosity Calculation

In this subsection, we will determine the luminosity for desired neutron output. The number of fusion events per second ($N_{DD}$) is given by the formula below:

$$N_{DD} = L \times \sigma \qquad (1)$$

Here, L presents the luminosity of the DD collider and $\sigma$ denotes the DD fusion cross-section. As can be seen from Table 2, the number of neutrons produced is half of the number of fusion events.

$$N_n = \frac{N_{DD}}{2} = L \times \frac{\sigma}{2} \qquad (2)$$

The relation between the output energy and luminosity can be expressed by the following formula:

$$L(cm^{-2}s^{-1}) = \frac{P(J.s^{-1})}{\sigma(cm^2) \times E(J)} \qquad (3)$$

Here, E represents the energy produced in a single fusion process. The number of neutrons produced can be expressed using the following formula (according to Eq. 2 and Eq. 3):

$$N_n = \frac{P}{2E} \qquad (4)$$

It is seen that $5.1 \times 10^{19}$ n/s are required to generate 10 MW power output solely from DD fusion.

The energy balance for the DD collider option is given in Table 3. For input electricity calculations, if 50 keV deuteron beams (100 keV for two beams) are used for a single neutron production the needed energy value became 600 keV: the factor of 3 arises from accelerator efficiency and the factor of 2 is due to the probability of neutron production in DD fusion (see Table 2).

Table 3. Energy Balance Q Values for *DD* Collider Fusions

|  | ① Muon catalyst cycle or cascaded neutron amplification | ② Fission Multiplication $k_{eff}$, M = 1/(1 - $k_{eff}$) | ③ Neutron flux | ④ Output Thermal (MW) | ⑤ Input Electric | ⑥ $Q_{economy}$ ④ ÷ ⑤ |
|---|---|---|---|---|---|---|
| *DD* collider | 1 |  | $5.1 \times 10^{19}$ | 10 | 4.90 MW | 2.04 |
| Cascade 1 | M1 = 20 | 0.95 | $2.55 \times 10^{18}$ | 10 | 245 kW | 40.8 |
| Cascade 1×Cascade 2 | M1×M2 = 1000 | 0.98, M2 = 50 | $5.10 \times 10^{16}$ | 10 | 4.90 kW | 2040 |
| Cascade 1×Cascade 2 main reactor | M1×M2×MR = 100 000 | 0.99, MR = 100 | $5.10 \times 10^{14}$ | 10 | 49 W | $2.04 \times 10^5$ |

The rearranged version of Eq. 2 for luminosity is:

$$L = 2 \times \frac{N_n}{\sigma} \qquad (5)$$

As can be seen from Figure 3, $\sigma$ is $3 \times 10^{-26} cm^2$ for 50 keV deuteron beams. According to Eq. 5, with $N_n$= 5.1×10[14] n/s, the luminosity value is calculated as $L = 3.4 \times 10^{40} cm^{-2}s^{-1}$.

### 3.1.2. Deuteron Accelerator Parameters

In this subsection, we will determine the accelerator parameters for the deuteron beam. The schematic illustration of deuteron accelerator is given in Figure 2.

The luminosity formula expressed in terms of the transverse size of deuteron beams ($\sigma_t$) is given below:

$$L = \left(\frac{N_D^2}{4\pi\sigma_t^2}\right) f_{rev} \qquad (6)$$

Here, $N_D$ represents the total number of particles in the deuteron accelerator ring and $f_{rev}$ is the revolution frequency of the deuteron beam. Under the assumption that the ring length is 10 m, the revolution frequency can be calculated as $f_{rev} = 2.2 \times 10^5 s^{-1}$ for 50 keV deuterons. According to Eq. 6, for $\sigma_t = 1\ mm$, the value of $N_D$ must be $1.4 \times 10^{17}$ to achieve $L = 3.4 \times 10^{40} cm^{-2}s^{-1}$ which is calculated in subsection 3.1.1.

Since 5.1×10[14] neutrons should be produced per second (see last row of Table 3), 7.2×10[14] deuterons per second will be used up for each beam. Therefore, to compensate for deuteron losses, the same number of deuterons must be injected into the rings per second. The deuteron source must therefore continuously supply the ring with a beam current of 0.115 mA. If the same source is used for initial filling, 194 seconds will be required to fill the ring. To fill the ring in 1 second, a source current of 22.4 mA is needed.

### 4. Conclusion

Hybrid reactors can make an exceptional contribution to meet humanity's energy needs and mitigate climate change. In this respect, the recently proposed combination of muon-catalyzed DD fusion with a cascade Thorium reactor is worthy of appreciation. It should be mentioned that Thorium reactors produce much less radioactive waste than Uranium reactors.

Our study demonstrates that a deuterium-deuterium (DD) collider can provide a viable and advantageous alternative to muon-catalyzed fusion in Thorium-based hybrid reactors. By utilizing two low-energy deuterium beams (50 keV) instead of the complex and costly 600 MeV proton accelerators required for muon production, the proposed system simplifies accelerator design and improves operational feasibility. The energy balance calculations reveal that the DD collider, in conjunction with cascade multiplication, significantly boosts system efficiency and scalability.

Future efforts should focus on experimental validation, beam dynamics optimization, and scaling the design for large scale energy generation to unlock the full potential of this innovative system.

Finally, the fixed target DD collision option can also be promising and our study on this subject is about to be completed. Let us note that due to large neutron yield, DD collision part of proposed system can be also used as a neutron generator.

**Acknowledgement**
The authors are grateful to A. Ozturk, B. Ketenoglu and G. Unel for useful discussions.